\documentclass[reqno]{amsart}
\usepackage{amsfonts}
\usepackage[active]{srcltx}
\usepackage{amsmath}
\usepackage{amssymb}
\usepackage{amsthm}
\usepackage{mathtools}
\usepackage{shuffle}
\usepackage{bm} 
\usepackage{upgreek}
\usepackage{graphicx} 
\usepackage{mathrsfs}
\usepackage{fancyhdr}
\usepackage{amsthm}
\usepackage{cases}
\usepackage{amsmath,amssymb,bm}
\usepackage{enumitem}
\usepackage[]{showlabels}
\usepackage{xcolor}

\usepackage{tikz-cd} 

\usepackage{enumitem}
\usepackage{wasysym}
\usepackage{mathscinet}
\usepackage{verbatim}

\usepackage{graphicx}
\usepackage[
bookmarks=true,         
bookmarksnumbered=true, 
colorlinks=true, pdfstartview=FitV, linkcolor=blue, citecolor=blue,
urlcolor=blue]{hyperref}
\usepackage{microtype}
\theoremstyle{plain}
\newtheorem{theorem}{Theorem}[section]

\newtheorem{problem}[theorem]{Problem}

\renewenvironment{proof}[1][Proof]{\textbf{#1.} }{\ \rule{0.5em}{0.5em} \par }
\theoremstyle{remark}

\theoremstyle{definition}

\let\Section=\section
\def\section{\setcounter{equation}{0}\Section}

\begin{document}


\title[Option Valuations of Futures Contracts with Negative Prices]{Practical Option Valuations of Futures Contracts with Negative Underlying Prices}
\author[]{Anatoliy Swishchuk} \address{Dept. of Mathematics \& Statistics,
University of Calgary,,
Calgary, Alberta, Canada T2N 1N4} \email{aswish@ucalgary.ca}

\author[]{Ana Roldan-Contreras} \address{Dept. of Mathematics \& Statistics,
University of Calgary,,
Calgary, Alberta, Canada T2N 1N4} \email{ana.roldancontreras@ucalgary.ca}
\author[]{Elham Soufiani} \address{ Department of Mathematics and Statistics,
University of Regina, SK} \email{esr434@uregina.ca}
\author[]{Guillermo Martinez} \address{} \email{gmomtzcanada@gmail.com}
\author[]{Mohsen Seifi} \address{Department of Statistics and
Actuarial Science (SAS)
Mathematics 3 (M3)
University of Waterloo} \email{smmseifi@uwaterloo.ca}
\author[]{Nishant Agrawal} \address{Dept of Mathematical and Statistical Sciences, University of Alberta at
Edmonton, Edmonton, Canada, T6G 2G1} \email{nagrawal@ualberta.ca}
\author[]{Yao Yao} \address{Dept. of Mathematics \& Statistics,
University of Calgary,
Calgary, Alberta, Canada T2N 1N4} \email{yao.yao1@ucalgary.ca}


\thanks{}

\begin{abstract}
 Here we propose two alternatives to Black 76 to value European option future contracts in which the underlying market prices can be negative or mean reverting. The two proposed models are Ornstein-Uhlenbeck (OU) and continuous time GARCH (generalized autoregressive conditionally heteroscedastic). We then analyse the values and compare them with Black 76, the most commonly used model, when the underlying market prices are positive.
\end{abstract}

\maketitle

\section{Introduction}
In March 2020, the prompt month WTI futures contract settled below zero for the first time in the contract’s history. Many market participants apply the Black 76 model or some variation when calculating the value of the options on this futures contract as a relatively straightforward, parametric valuation method. This calculation model is hard wired into many Commodity Trading and Risk Management Systems. Traders and risk managers rely on its straightforward and reproducible output.

However, Black 76 requires positive underlying market prices. The negative prompt month settlement price caused considerable consternation among energy traders and risk managers.

More generally, OTC options are also available on basis or differential prices. These transactions are options on the difference between two published indexes such as NYMEX Henry Hub and AECO (for natural gas) or Cushing WTI and Houston (for crude oil). As such, these instruments frequently have negative underlying market prices.

Our task is to propose alternative models to Black 76 to valuate option prices when the underlying future contracts can assume negative values.

Our methodology is the following one: 
\begin{enumerate}
    \item Take data (prices), sketch their behaviour, i.e., their evolution in time;
    \item If the prices are positive and not mean-reverting, then use geometric Brownian motion (GBM) model for their evolution and Black-76 (\cite{Bl}) formula for option valuation of futures (see also formulas (BlCall) and (BlPut) in \cite{SwishchukLectureNotes})
    \item If the prices are positive and mean-reverting, then use continuous-time GARCH (or, another name, inhomogeneous GBM model) model \cite{Swishchuk} and option pricing formula (35) from \cite{Swishchuk}, Theorem 5.1;
    \item If the prices are both positive and negative, but not mean-reverting, then use Bachelier model and his formula, \cite{Ba} (see also formulas (Ba\_1) and (Ba\_2) in \cite{SwishchukLectureNotes});
    \item If the prices are both positive and negative, and mean-reverting with mean-reverting level $0,$ then use Ornstein-Uhlenbeck model \cite{UO} and the formulas (OUCall\_1) and (OUCall\_2) from \cite{SwishchukLectureNotes};
    \item If the prices are both positive and negative, and mean-reverting with mean-reverting level non-zero, then use Vasicek model \cite{Vasicek} and the formulas (VasCall\_1) and (VasCall\_2) from \cite{SwishchukLectureNotes}.

We note, that the most general, to the best of our knowledge, stochastic models for spot prices (both arithmetic and multiplicative) in electricity and related markets were presented in \cite{Be}.
\end{enumerate}

In this paper we show how this methodology works on data sets presented by Scott Dalton (Ovintiv Services Inc.), namely, we use WTI data set and NYMEX NG data set.

\section{Definitions} 

A \textbf{primary security} (or securities for short) is any asset that can be traded independently from any other asset, such as stocks. A \textbf{derivative security} or (or derivatives) are legal contracts conferring financial rights or obligations upon the holder.

A \textbf{forward contract} is an agreement to buy or sell a risky asset (such as crude oil or natural gas) at a determined future date $T,$ known as \textbf{delivery date}, at a specified price $K,$ known as \textbf{delivery price}. The price of the asset (or commodity) at time $t$ is known as \textbf{forward price} and denoted by $F(t, T).$ Notice $K = F(0, T).$

Similarly, a \textbf{future contract} (or futures for short) involves an underlying asset, which we typically take as a forward contract, and a specified \textbf{delivery date} $T.$ A future price set at time $t$ with delivery date $T$ will be denoted as $f(t, T).$

An \textbf{European option} is a derivative security contract that gives the holder the right, but not the obligation to buy or sell the underlying asset, for a price $K$ fixed in advance, known as \textbf{exercise} or \textbf{strike} price, at a specified future time $T_e,$ known as \textbf{exercise} or \textbf{expiry date.} An option contract with expiry date $T_e$ stops being valid after this time. The option is known as a \textbf{call option} if the holder has the right to \emph{buy} the asset, while a \textbf{put option} gives the holder the right to \emph{sell} the asset.

Forwards and futures are legal agreements between two parties giving obligations between them, in contrast, options are legal agreements giving rights to the holder. Because of this advantage intrinsic in options (the holder may trigger the contract should it be in their favour) is that they are to be purchased. We are concerned is valuing them, specifically, we are interested in valuing European call options for futures prices.





\section{Proposed alternative models to Black 76.}
Black 76 model is obtained from the more general Black-Scholes model (1973). Black-Scholes is a model for the price of a stock at time $t$ and it is given by the following Stochastic Differential Equation (SDE)
\begin{displaymath}
    dS_t = \mu S_t dt + \sigma S_t dW_t,
\end{displaymath}
where $0 \leq t \leq T$ represents time ($T$ is the expiry date), $\mu \in \mathbf{R}$ is a number known as the ``drift'', $\sigma > 0$ is the ``volatility'' and $(W_t)_{t \geq 0}$ is Wiener process (or Brownian motion). In this model, $S_0$ is deterministic (not random) and known in advance. Using It\={o}'s formula, it can be deduced that
\begin{displaymath}
    S_t = S_0e^{(\mu - \frac{1}{2} \sigma^2)t + \sigma W_t}.
\end{displaymath}
This shows that under the assumptions of Black-Scholes, the stock price will be positive (assuming $S_0 > 0$) for all times.

    \subsection{Orsntein-Uhlenbeck (Vasicek) model (1930/1977)}
    The first alternative we propose to Black 76 is given by the following Orsntein-Uhlenbeck SDE
    
    \begin{eqnarray}
        dS_t &=& a(b - S_t) dt + \sigma dW_t, \label{Eq: OU SDE}
    \end{eqnarray}
    
    where $a, \sigma > 0$ and $b \in \mathbf{R}.$ Here $a$ is known as the ``reversion rate'', $b$ as the mean and $\sigma$ as the volatility. Again, using It\={o}'s formula, it can be shown that the solution to the OU SDE is given by
    \begin{equation}\label{Eq: OU SDE solution}
        S_t = e^{-at} S_0 + b(1 - e^{-at}) + \sigma e^{-at} \int\limits_0^t e^{as} dW_s.
    \end{equation}
    This is a Gaussian random variable with mean $e^{-at} S_0 + b(1 - e^{-at})$ and variance $\sigma^2(1 - e^{-at})/2a.$ It is readily seen it can assume negative values and as $t \to \infty,$ this Gaussian random variable converges in distribution to a Gaussian with mean $b$ and variance $\sigma^2/2a,$ the rate of convergence is given by $a.$ The value of an European call option at time $t$ with delivery date $T_e,$ rate of risk-free investment $r,$ and strike price $K$ is given according to (see formulas (VasCall\_2) in \cite{SwishchukLectureNotes})
    \begin{equation}\label{Eq: Euro Call Opt Price for OU}
        C(F, T_e) = e^{-r(T_e - t)} \left[ \xi_+(t, T_e)  \Phi \left(\dfrac{\xi_-(t, T_e)}{\zeta} \right) + \zeta \Phi' \left(\dfrac{\xi_-(t, T_e)}{\zeta} \right) \right],
    \end{equation}
    in which $\Phi$ is the distribution function of a standard Gaussian random variable and
    \begin{align*}
        \xi_\pm(t, T_e) &= e^{\pm aT_e} (F(t, T_e) - b) - K \\
        \zeta &= \sigma \sqrt{\dfrac{1-e^{-2a T_e}}{2a}}
    \end{align*}
    The future prices of this model will be modelled using (\ref{Eq: OU SDE}).

    \subsection{Continuous Time GARCH model:}
    Some times the commodity prices exhibit different behavior with respect to time, which is known as Mean-Reversion. It means that, unlike stock prices that tend to change around zero, they tend to return to a non-zero long-term mean. Therefore for a risky asset $S_t$ which has a mean reverting stochastic process, we have the following SDE:
    \begin{equation}\label{Eq: C-T GARCH}
     dS_t = a(b - S_t)dt + \sigma S_t dW_t 
    \end{equation}
    where W is a standard wiener process, $\sigma >0$ is the volatility, the constant $b \in \mathbb{R}$ is the mean reversion level (the long term mean), and $a>0$ measures the rate (or the strength) of our mean reversion. The closed form of the above equation for a European Call has been provided in section (4).

\section{Methodology and results}
\subsection{OU model}
    According to (\ref{Eq: OU SDE}), we need to calibrate the parameters $a,$ $b$ and $\sigma.$ Using (\ref{Eq: OU SDE solution}) (in which $S$ is substituted for the future price $F$) it can be seen that observations of the future price are in a linear relation plus normally distributed error terms. As such, least-squares linear regression can be used. In Fig. \ref{fig:NatGas} we do the calibration of the parameters for the OU model using Natural Gas future prices provided by Ovintiv. We can see the prices are around the mean, which is an assumption of validity of the model.
    
\subsection{WTI Dataset}
    For WTI crude oil futures, we compare the option prices calculated by the Black-76 model and the Vasicek model.
    Let $C(t, T_e)$ be the value for the European call option written on a forward F. Then the Black-76 formula for European call option price is:
    \begin{equation}
        C(t, T_e)=e^{-r(T_e-t)}[F(t,T)N(d_1)-KN(d_2)],
    \end{equation}
    where $d_{1,2}:=\frac{ln(F/K)\pm\frac{1}{2}\sigma^2(T_e-t)}{\sigma\sqrt{T_e-t}}$.

    The European call option formula for Vasicek model is similar to equation (3)
    with slight differences:
     \begin{equation}\label{Eq: Euro Call Opt Price for OU}
        C(F, T_e) = e^{-r(T_e - t)} \left[ \xi_+^*(t, T_e)  \Phi \left(\dfrac{\xi_-^*(t, T_e)}{\zeta} \right) + \zeta \Phi' \left(\dfrac{\xi_-^*(t, T_e)}{\zeta} \right) \right]
    \end{equation}
    in which $\Phi$ is the distribution function of a standard Gaussian random variable and
    \begin{align*}
        \xi_\pm^*(t, T_e) &= e^{\pm aT_e} (F(t, T_e) - b^*) - K \\
        \zeta &= \sigma \sqrt{\dfrac{1-e^{-2a T_e}}{2a}}
    \end{align*} 
    
where $b^{\ast}=b-\lambda\sigma/a, \lambda\in\mathbb{R}$ is a market price of risk.

\begin{figure}[h]
        \centering
        \includegraphics[scale = 0.6]{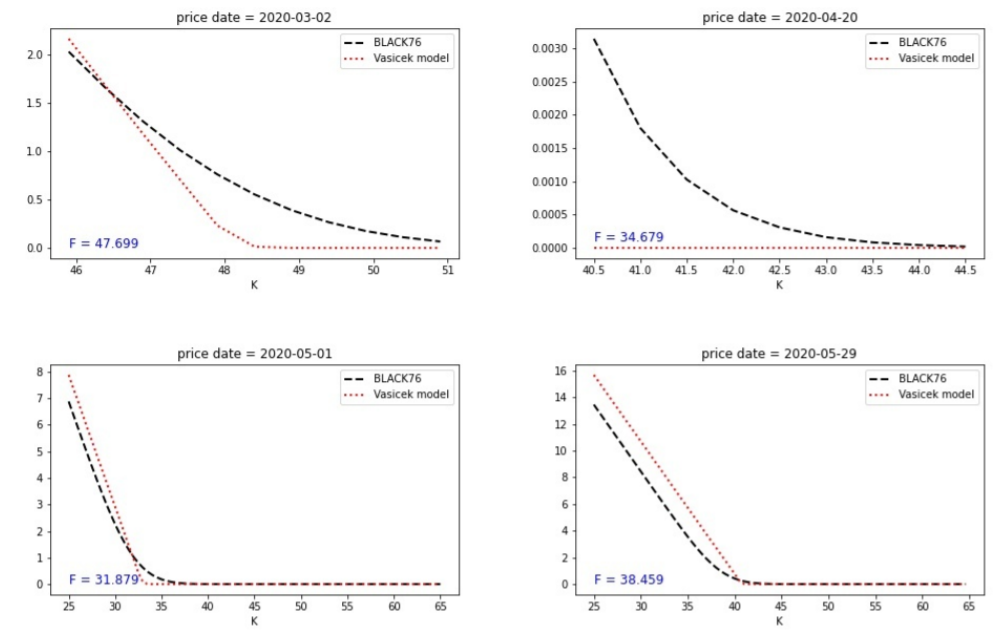}
        \caption{Black76 vs Vasicek models for Call option prices}
        \label{fig:WTI}
    \end{figure}

\begin{figure}[h]
        \centering
        \includegraphics[scale = 0.3]{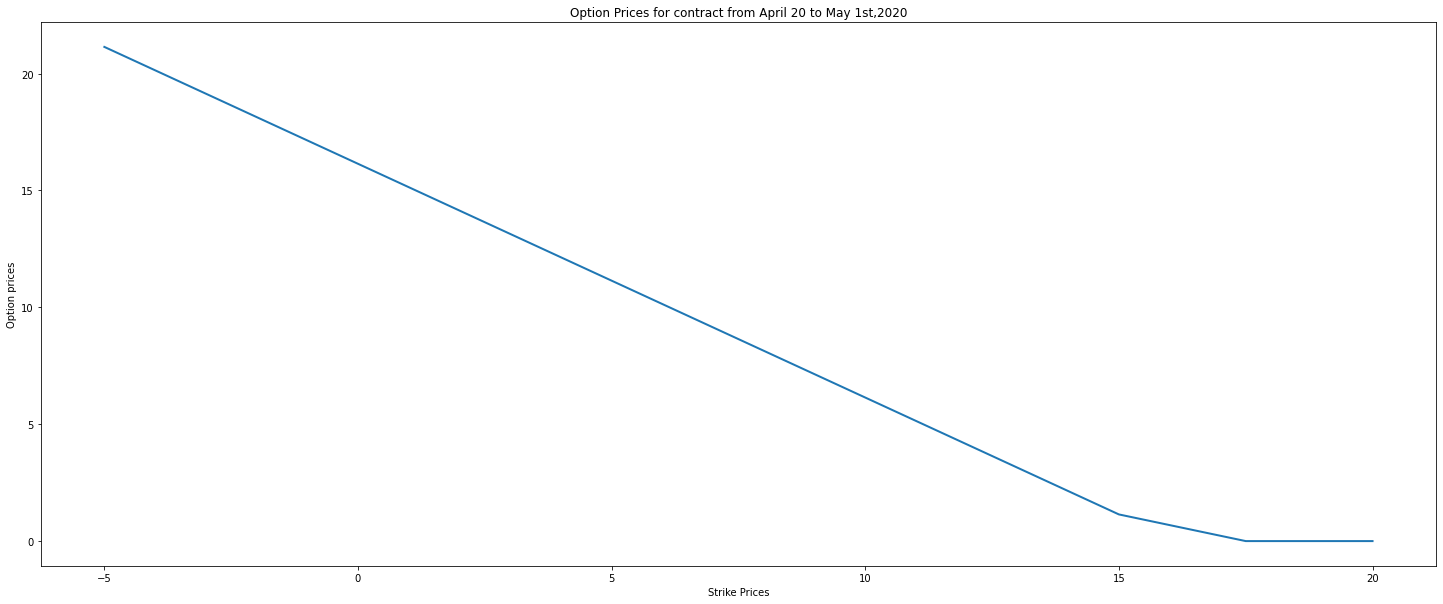}
        \caption{Option prices if futures prices becomes negative}
        \label{fig:WTI2}
    \end{figure}

We use the above formula for black 76 and monte carlo simulation to get the graphs \ref{fig:WTI} of option prices. Each graph in Figure \ref{fig:WTI}  shows the option prices with different strike price K. Except for the chart with the price date of 2020-04-20, when we have a negative future price, the others are all positive. From the graphs we can see that option prices on futures computed by the Vasicek model are very close to the prices calculated by the Black-76 model when future prices are positive.
When future prices are negative we can employ Vasicek model again to come up with the call option prices. Here Black 76 model would fail as it does not accepts the negative prices. Figure \ref{fig:WTI2} shows the price of the option for various strike prices where we were also able to calculate prices for negative strike prices. These prices have been calculated using Monte Carlo simulation.

\subsection{NYMEX Natural Gas Dataset} 
\begin{figure}[h]
        \centering
        \includegraphics[scale = 0.32]{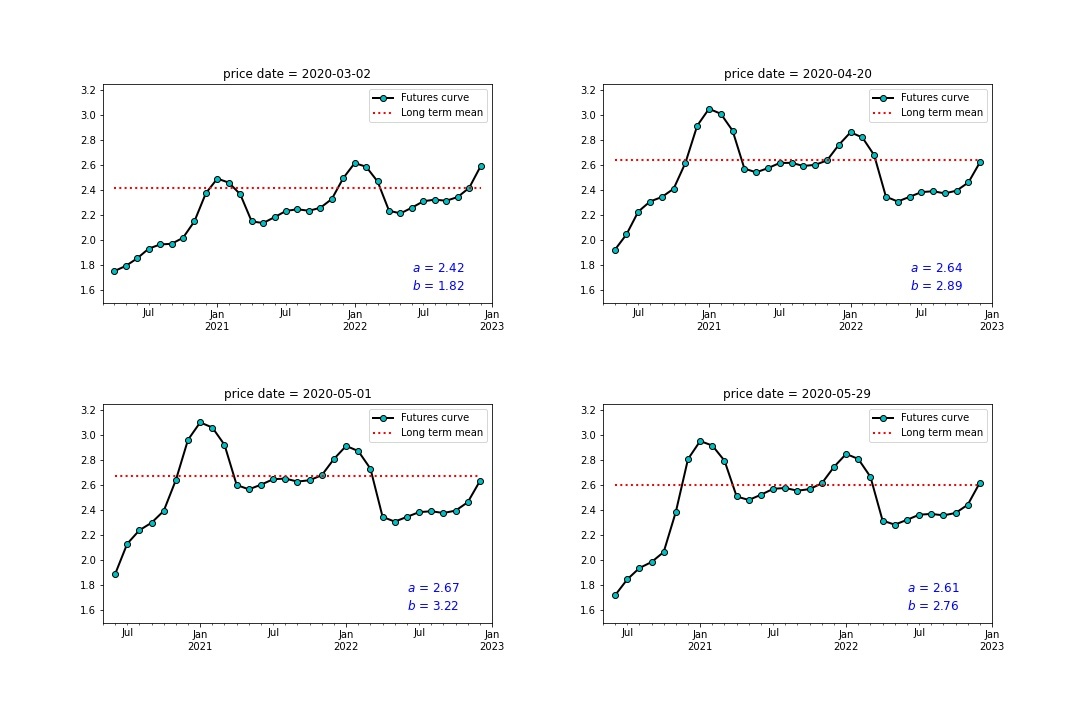}
        \caption{Calibrations of parameters for different initial ``price dates.'' Here the $x$-axis is the expiry date ($T$) and the $y$-axis is the price per unit. The dotted line is the mean value $b.$}
        \label{fig:NatGas}
    \end{figure}

    For the Natural Gas (NG) dataset, we will see that all the underlying future prices are positive. However, they exhibit a non-zero mean-reversion process over the time (figure \ref{fig:NatGas}).
    In this case, for the corresponding option prices, we used the Continuous-Time GARCH model (as in equation \ref{Eq: C-T GARCH}).

    In this methodology, first we should consider the model (\ref{Eq: C-T GARCH}) under a risk-neutral probability $P^*$. Therefore, in a risk-neutral world our model will take the following look (for more details see \cite{Swishchuk}, sec. 5.4):
    \begin{equation} \label{Eq: C_T GARCH risk-neutral}
        dS_t = a^* (b^* - S_t)dt + \sigma S_t dW^*_t
    \end{equation}
    where:
    \begin{equation*}
        a^* := a + \lambda \sigma ,  b_*:= \frac{ab}{a+\lambda \sigma}
    \end{equation*}
    and $W^*_t$ is defined as 
    \begin{equation*}
        W^*_t := W_t + \lambda \int _{0}^{t} S(u) du.
    \end{equation*}
    Here, $\lambda \in \mathbb{R}$ is the \textit{market price of risk}.\\
    For this model (\ref{Eq: C_T GARCH risk-neutral}) we have an explicit option pricing formula for European Call option \cite{Swishchuk}: 
    \begin{align*}
        C_{T}^{*} = & e^{-(r+a^*)T} S(0) \Phi(y_+) - e^{-rT} K \Phi(y_-) + \\ 
        & b^* e^{-(r+a^*)T} \left[(e^{a^* T}-1) - \int _{0}^{y_0} z F_{T}^{*}(dz)] \right]
    \end{align*}
    where, $y_0$ is the solution of:
    \begin{align*}
        y_0 = & \tfrac{\ln{\tfrac{K}{S(0)} + \left(\tfrac{\sigma ^2}{2}+a^*\right)T}}{\sigma \sqrt{T}} - \\
        & \tfrac{\ln{\left(1 + \tfrac{a^* b^*}{S(0)}\right)} \int _{0}^{T} e^{a^*s}e^{-\sigma y_0 \sqrt{s} + \tfrac{\sigma^2 s}{2}}ds}{\sigma \sqrt{T}}
    \end{align*}
    with,
    \begin{equation*}
        y_+ := \sigma \sqrt{T} - y_0, \quad \textrm{and} \quad y_- := -y_0,
    \end{equation*}
    and, $F_{T}^{*}(dz)$ is the probability distribution under the risk-neutral probability $P^*$, as in \cite{Swishchuk}. 
    \subsubsection{Methodology and results}
    In this approach, to avoid the huge computations regards to the explicit formula, we used Least Square Regression method for calibrating the parameters by following the methodology in \cite{CalibratingOU},
    \begin{equation*}
        F_{i+1} = \tau F_{i} + \mu + sd(e),
    \end{equation*}
    to have the following equations:
    \begin{align*}
        & F_x = \sum _{i=1}^{n} F_{i-1}, 
        & F_y = \sum _{i=1}^{n} F_{i}, \\
        & F_{xx} = \sum _{i=1}^{n} F_{i-1}^2, 
        & F_{yy} = \sum _{i=1}^{n} F_{i}^2, \\
        & F_{xy} = \sum _{i=1}^{n} F_{i-1}F{i}
    \end{align*}
    and then the following relationships can be considered:
    \begin{align*}
    \centering
        & \tau = \frac{n F_{xy} - F_x F_y}{n F_{xx} - F_{x}^{2}}, \\
        & \mu = \frac{F_y - \tau F_x}{n}, \\
        & sd(e) = \sqrt{\frac{nF_{yy} - F_{y}^{2} - \tau(n F_{xy} - F_x F_y)}{n(n-2)}}.
    \end{align*}
    For our purpose, we used the Euler approximation to simulate the future prices in order to approximate the corresponding European Call option prices.
    \begin{equation}
        F_{i+1} = F_{i} \exp{a^* \delta} + b^* (1- \exp{-a^* \delta}) + \sigma F_{i} \sqrt{\frac{1-\exp{-2 a^* \delta}}{2 a^*}} N_{0,1}
    \end{equation}
    Here, $\delta > 0 $ is a time space, and the $F_{i}$ prices are the exact discrete solution of equation (\ref{Eq: C-T GARCH}). Hence, we can find the following relations between the parameters:
    \begin{equation*}
        a = - \frac{\ln{\tau}}{\delta} , \\
        b = \frac{\mu}{1 - \tau},\\
        \sigma = sd(e) \sqrt{\frac{- 2 \ln{\tau}}{\delta (1-\tau ^2)}}
    \end{equation*}
    Finally, for the risk neutral parameters, the following adjustment has been applied:
    \begin{equation*}
        a^* = a + \lambda \sigma,   b^* = \frac{a b}{a + \lambda \sigma}
    \end{equation*}
    According to our dataset,  there was not any access to the market option prices to estimate the market price of risk. Therefore, the following formula has been taken into account:
    \begin{equation*}
        \lambda := \frac{\tfrac{dF}{F} - r}{\sigma}
    \end{equation*}
    where, $\frac{dF}{F}$ is a returns on futures prices, $r$ is the interest rates, and $\sigma$ is the implied volatilities.
    \begin{figure}
        \centering
        \includegraphics[scale = 0.5]{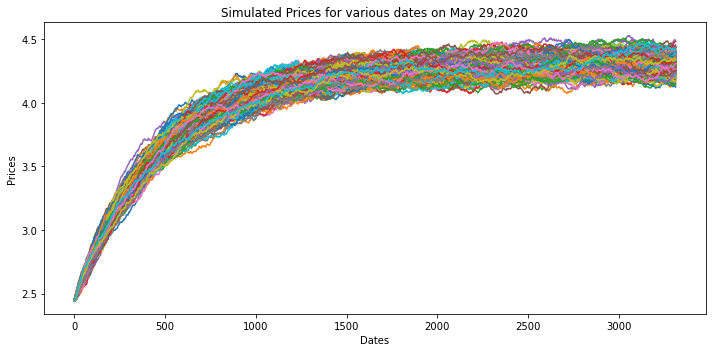}
        \caption{The evolution of simulated future price with respect to time}
        \label{fig:sim_NG}
    \end{figure}
   
\begin{figure}
        \centering
        \includegraphics[scale = 0.5]{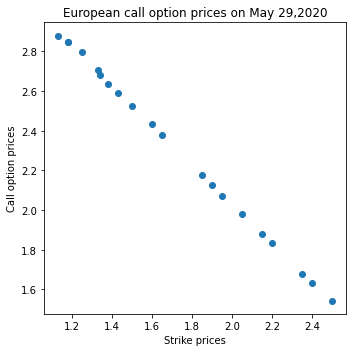}
        \caption{In this picture, we can see the evolution of the calculated option prices, according to the Continuous-Time GARCH model, with respect to their related strike prices is depicted}
        \label{fig:GARCH_option}
    \end{figure}

\begin{figure}
        \centering
        \includegraphics[scale = 0.9]{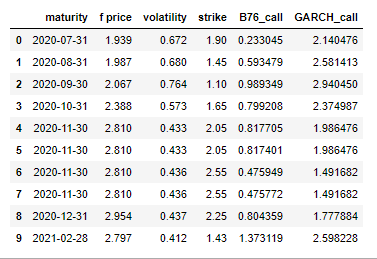}
        \caption{In this table, the accuracy of our model comparing to the known Black-76 model has been exhibited for the first 10 strike prices}
        \label{fig:GARCH_B76_calloption}
    \end{figure}

    The future prices were simulated 20 times (an exercise of this is shown in figure \ref{fig:sim_NG}), and the average of them is applied in the payoff function. Then, the discount of the average of payoffs considered as the requested call option prices with continuous-time GARCH model approach (results can be seen through figure (\ref{fig:GARCH_option}) and (\ref{fig:GARCH_B76_calloption})).
    
    Here, the risk-neutral parameters $a^*$ and $b^*$ has been estimated as $1.68528518$ and $2.64820985$ respectively. 
    
 \section{Conclusion}
In this project, we worked with some useful alternative models which are helpful for valuation of options on future contracts. In spite of Black 76 is the most commonly used model for valuating option future contracts in industry, it is necessary to have alternative models for the valuation when the prices' behaviour differs from the prices describe by the same model. For WTI future option prices, O-U and Vasicek model has shown to have similar prices as Black 76 when future prices are positive and have a valuation when negative prices, which is useful when irregular events happen. Also, for Natural Gas future option prices, continuous time GARCH also display comparable values as Black 76, further it allows us to calibrate a mean reversion parameter to describe in a better way future option prices and their behaviour.

As a recommendation, it would be useful for industry to keep a track on this two models to know how to react in unusual situations and double check their own valuation prices. This models have shown to be simple to understand, clear to calculate and comparable with what the industry uses. With respect to the data and results, in table \ref{table:recommendation} are some suggestions for the valuation according to the data's nature:
  
\begin{table}[!h]
\begin{center}
\begin{tabular}{l l l}
\hline
\textbf{Future Prices} & \textbf{Mean-Reversion Level} & \textbf{Model}\\
\hline
Positive & none
 & GBM model \\
Positive & b
 & Continuous-time GARCH \\
Negative and Positive & 0
 & OU model \\
Negative and Positive & b
 & Vasicek model \\
 Negative and Positive & none & Bachelier model\\
\hline
\end{tabular}
\end{center}
\caption{Recommended model according to sign of prices and mean reversion behaviour.}
\label{table:recommendation}
\end{table}

\section*{Acknowledgement}

We thank Scott Dalton (Ovintiv Services Inc.) for his time, for providing the datasets, and his willingness to share his industry expertise. We also thank PIMS committee and Professor Kristine Bauer for well-organized workshop. 



\end{document}